\documentclass[%
 reprint,
 amsmath,amssymb,
 aps,
]{revtex4-1}

\usepackage{graphicx}
\usepackage{dcolumn}
\usepackage{bm}

\usepackage{color}

\newcommand\nc{\newcommand}
\nc\be{\begin{eqnarray}} \nc\ee{\end{eqnarray}}
\nc\bes{\begin{eqnarray*}} \nc\ees{\end{eqnarray*}} \nc\pa{\partial}
\nc\pad[2]{\frac{\pa #1}{\pa #2}} \nc\padd[2]{\frac{\pa^2 #1}{\pa
{#2}^2}} \nc\nd[2]{\frac{d #1}{d #2}} \nc\pat[2]{\frac{D #1}{D #2}}
\nc\ov{\overline} \nc\degree{^{\circ}} \nc\ord[1]{{\cal O}(#1)}
\nc\ra{\rightarrow} \nc\Ra{\Rightarrow} \nc\dint{{\mbox ~ d}}
  \nc\dg{{\dot \gamma}}

\newcommand{\beas}{\begin{eqnarray*}}
\newcommand{\eeas}{\end{eqnarray*}}
\newcommand{\bi}{\begin{itemize}}
\newcommand{\ei}{\end{itemize}}

\newcommand{\lb}{\label}

\newcommand{\bea}{\begin{eqnarray}}
\newcommand{\eea}{\end{eqnarray}}

\begin{document}


\title{Optical diffraction from isolated nanoparticles\\
}%

\author{T.G. Myers}
\email{tmyers@crm.cat}
\affiliation{
 Centre de Recerca Matem\`{a}tica\\ Bellaterra, Spain
}%

\author{H. Ribera and W.S. Bacsa}
 \email{hribera@crm.cat, wolfgang.bacsa@cemes.fr}
\affiliation{Centre d'Elaboration de Materiaux et d'Etudes Structurales and University of Toulouse\\ Toulouse,
 France
}%

\date{\today}

\begin{abstract}

When subjected to monochromatic incident light a nanoparticle will emit light which then interferes with the incident beam. With sufficient contrast and sufficiently close to the particle this interference pattern  may be recorded with a pointed optical fiber in collection mode. It is shown that the analytic dipole model accurately reproduces the observed interference pattern.
Using this model and measuring only the lengths of the first two major axes of the observed elliptical fringes we are able to
reproduce and quantify the fringe pattern. Importantly, we are able to locate the nanoparticle, with respect to the fibre, using only visible light in a simple experimental setup. For the case described where the image plane is of the order microns above the substrate, hence the fringe number is large, it is shown that the prediction for the particle location and fringe number is insensitive to measurement errors. The phase shift of the scattered wave, a quantity that is notoriously difficult to measure, is easily determined from the theory however it is very sensitive to errors.
\end{abstract}

\pacs{Valid PACS appear here}
\maketitle



Due to quantum size effects and their large surface to volume ratio nanoparticles have very distinct properties when compared to their macroscale counterparts. These properties may be exploited by adding nanoparticles to materials, which then also exhibit altered behaviour. The extent of the alteration depends on the concentration and distribution of  particles. Consequently, when developing a nanoparticle infused material the ability to identify the particle distribution is crucial. Similarly other processes, such as the mass production of nanoparticles by industrial methods, requires a detailed screening process.

The observation of single fluorescence label free nanoparticles typically requires the use of transmission or scanning electron microscopy. The use of  electron beams ($>$ 50 kV acceleration voltage) requires a  vacuum or low pressure environment which often affects the particle's properties and changes its structure. It is therefore desirable to be able to use a less invasive optical method within an ambient atmosphere. In the past this has been impeded by the difference between the wavelength of visible light and the particle size, meaning that the smallest label free particles that may be observed in this manner are of the order 200 nm (around half the optical wavelength). Optical observation is therefore usually only applied to ensembles of nanoparticles and often in the far field \cite{Schatz}. To date, even though isolated nanoparticles can be observed in near field optics, quantitative optical measurements on individual nanoparticles  has been deemed impossible \cite{Fillard}. The optical response is drastically reduced for nanoparticles due to their small size. Optical resonances enhance the optical response for a particular wavelength.  For example, in metal particles plasmon resonances may be used \cite{Kar}.

In this letter we will briefly describe an experiment using optical interference and a simple reflection geometry which allows us to accurately locate individual nanoparticles using visible light. A theory is developed which describes the interference pattern. Combining theory with simple measurements of the observed  pattern we are able to identify the particle location. The phase shift of the scattered wave is also readily deduced from the result.

\begin{figure}[h!]
\vspace{0.5cm}
	\begin{center}
	\includegraphics[width=0.4\textwidth]{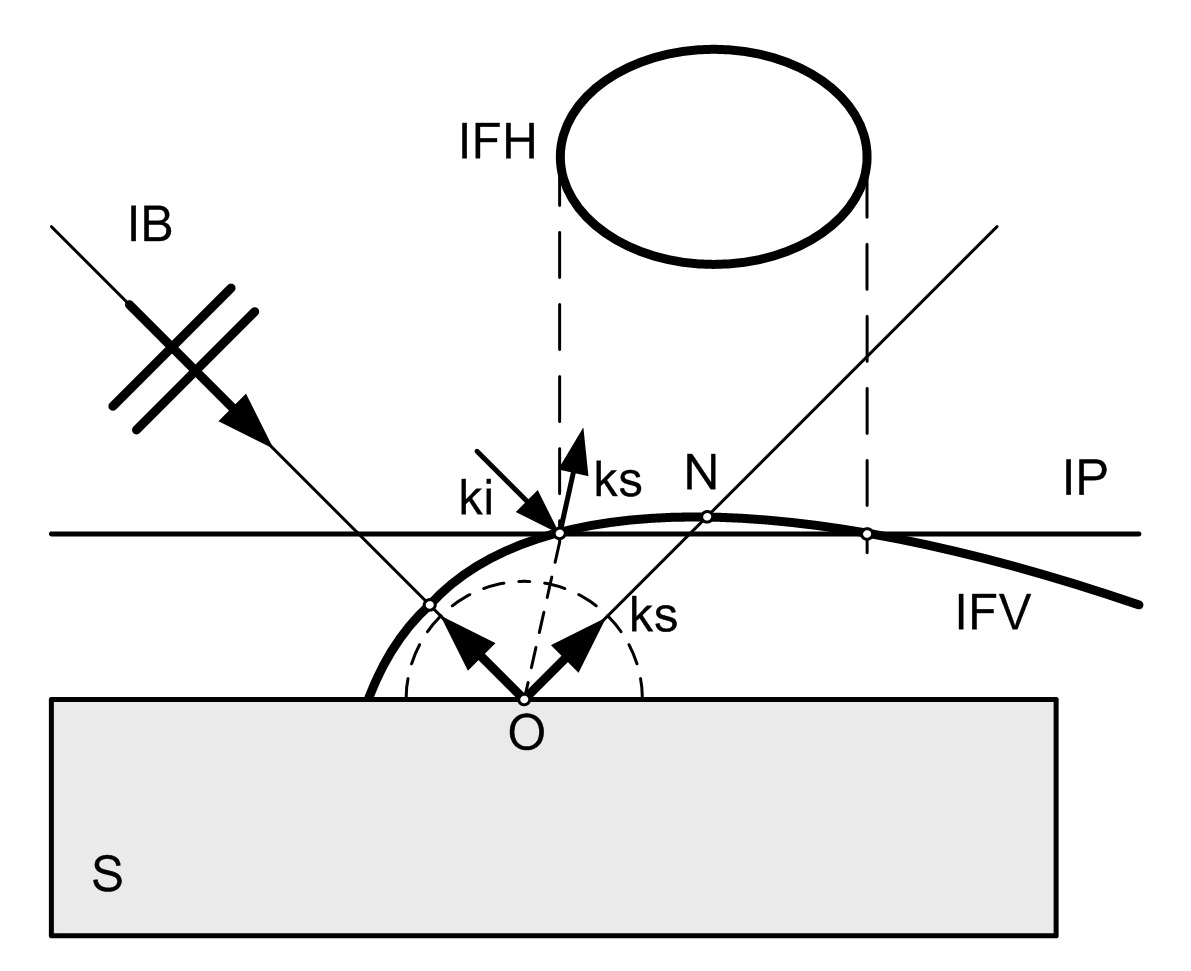}\vspace{-2.0cm}
	\end{center}	
  \vspace{1.8cm}
	\caption{Schematic of geometry considered: incident beam (IB), scatterer (O), substrate (S), image plane (IP), Vertical Interference Fringe (IFV, which has a parabolic form) and horizontal interference fringe (IFH, elliptical form). For simplicity only one interference fringe is shown. ki and ks are the wave vectors of the incident and scattered wave. The line through ON shows the reflected beam direction. N is the maximum of the parabolic interference fringe in the vertical plane. Dashed line indicates the angular range of ks. }
	\label{Setup}
\end{figure}
In the experiment nanoparticles are placed on an opaque substrate and illuminated with a laser beam at a known incident angle. The overlapping incident and reflected beam form surface standing waves parallel to the surface. The particle causes light to be scattered.
The coherence of the incident and scattered fields produces an interference pattern with an intensity proportional to the incident field. So, even though the scattered wave is small in amplitude its effect on the incident field results in observable lateral standing waves.
Both type of standing waves are observed by scanning a pointed optical fiber in collection mode \cite{Bacsa}. The surface standing waves are used to orient the image plane to ensure  that it is parallel to the substrate, in this manner we are able to scan at a variable distance from the surface and record the lateral standing waves. The experimental set-up is presented in Figure \ref{Setup}.

\begin{figure}[htb!]
	\begin{center}
	\vspace*{-2.0cm}
	\includegraphics[width=0.5\textwidth]{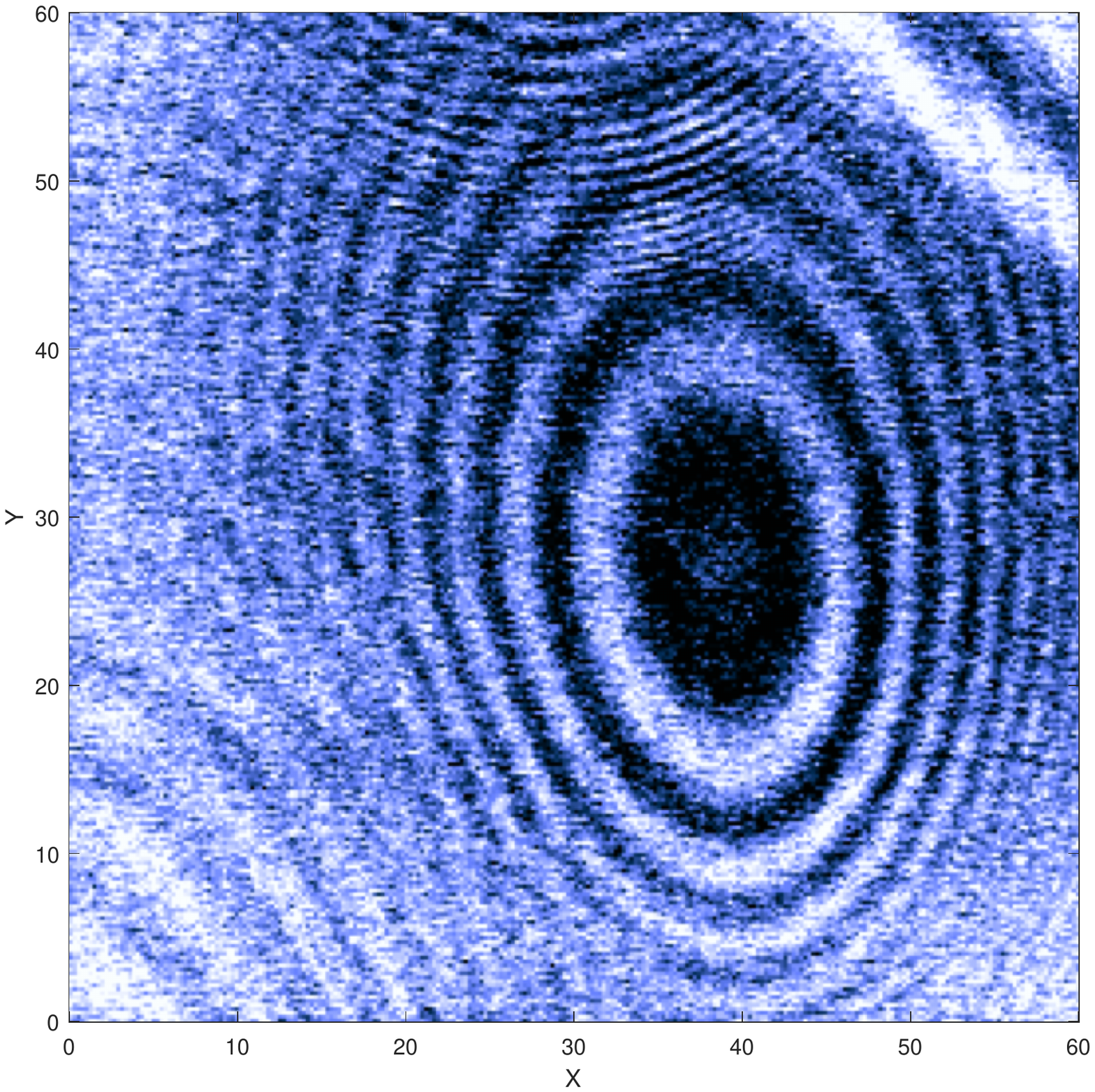}\vspace{-3.0cm}
	\end{center}	
	\caption{Typical interference pattern caused by the interaction between the incident field and the field scattered by a nanoparticle. }
	\label{BigFringe}
\end{figure}

We now analyze the horizontal interference fringes (parallel to the substrate) formed by the lateral standing waves.  Provided the incident beam is not perpendicular to the substrate the interference pattern due to a single nanoparticle is a series of ellipses. (The pattern in a vertical plane is a series of parabolas.) A typical result is presented in Figure \ref{BigFringe}. It shows the horizontal interference pattern produced by illuminating a surface with  an incident beam of wavelength 632 nm at an angle of approximately 53$^{\circ}$ (to the vertical). The image size is 60 $\times$ 60 $\mu$m. In this case we do not know the height of the image plane.
\vspace{-0.5cm}

The theoretical model is based on the observation that, while the light scattered by the particle cannot be observed using the optical fiber its effect on the incident and reflected beam can. When describing the scattered wave by an electric dipole and the incident beam with a plane wave, the intensity variation caused by the interaction is proportional to
\bea
\label{Iceq}
I_c =  \cos \left((\mathbf{k}_s - \mathbf{k}_i) \cdot \mathbf{r} - \psi \right) ~ ,
\eea
where $\mathbf{k}_i= k(0, \sin \theta_i, -\cos \theta_i)$, $\mathbf{k}_s = k \hat{\mathbf{r}}$, $k = (2\pi)/\lambda$ is the wavenumber, $\lambda$ the wavelength of the incident beam, $\theta_i$ the angle of incidence and  $\psi$ the phase shift, see \cite{Klein}. The co-ordinate system is defined such that the particle lies at the origin. To avoid mixing spherical and Cartesian systems we write $\mathbf{r} = (x,y,z) = r (\sin \theta \cos \phi, \sin \theta \sin \phi, \cos \theta)$, where $r = |\mathbf{r}| = \sqrt{x^2+y^2+z^2}$.  The Cartesian system $\mathbf{x}=(x,y,z)$ is chosen such that the substrate is the plane $z=0$ and $y$ is aligned with the major axis of the first ellipse.

Maxima of fringes occur when the argument of equation \eqref{Iceq} is some integer multiple of $2 \pi$. After substituting for
 $\mathbf{k}$ and $\hat{\mathbf{r}}$ and rearranging the surfaces of maximum brightness may be defined by
\bea
\sqrt{x^2+y^2+z^2} - y \sin \theta_i  + z \cos \theta_i    =  \left( n  + \frac{\psi}{2\pi}\right) \lambda , ~ \lb{parabeq}
\eea
for integer $n$. Since initially the position of the particle is unknown we introduce a second co-ordinate system $\mathbf{X}=(X,Y,Z)$ such that the origin is located at the bottom left corner of the experimental image of Figure \ref{BigFringe}, with $X$ horizontal, $Y$ vertical and $Z$ equivalent to $z$.

In this example the image of Figure \ref{BigFringe} was taken at an unknown height, which we denote $z=Z=z_c$. Using equation \eqref{parabeq} we may describe the recorded interference pattern by
\bea
\sqrt{x^2+y^2+z_c^2} - y \sin \theta_i     =  \mu_n - z_c \cos \theta_i = \Lambda_n, ~ \lb{parabeq2}
\eea
where $\mu_n =  \left( n  + \frac{\psi}{2\pi}\right) \lambda$. This can be rearranged to give the equation for an ellipse
\bea
\label{Ellipseeq}
\frac{x^2}{a_n^2} + \frac{(y-y_{cn})^2}{b_n^2} = 1
\eea
where the centre of the ellipse is located at $(0,y_{cn})$ and
\bea
a_n^2 = b_n^2 \cos^2 \theta_i ~,\qquad b_n^2 = \frac{\Lambda_n^2-z_c^2\cos^2 \theta_i}{\cos^4\theta_i} \, . \nonumber
\eea
From equations (\ref{parabeq2},\ref{Ellipseeq}) we can determine necessary information to characterise the interference pattern. First we note that the ratio of axes, $a_n/b_n = \cos \theta_i$, depends only on the angle of incidence. This provides a simple check on either the measurements of the axes or the angle of incidence. Second, the centre of each fringe is located at
\bea
\label{ycneq}
y_{cn} = \frac{\Lambda_n \sin \theta_i}{\cos^2\theta_i} ~ .
\eea
The distance between two consecutive centres is then
\bea
y_{c(n+1)}-y_{cn} = \frac{\lambda \sin \theta_i}{\cos^2\theta_i}
\eea
which depends only on the angle of incidence and the wavelength.

Now consider the expression for $b_n$ which contains unknowns $z_c, n, \psi$. The final two only occur in the combination specified by $\mu_n$, so instead we may deal with only  two unknowns $z_c, \mu_n$. Two experimental measurements $b_n, b_m$ will then be sufficient to determine the unknowns. Further, noting that $\psi \in [0,2\pi)$, once the value of $\mu_n$ is found we may write it in the form $\mu_n = (n+\psi/(2\pi))\lambda= C \lambda$ and then $n$ is the integer part of $C$ and the remainder  is $\psi/(2 \pi)$. Thus we determine three unknown quantities from only two measurements. With $z, n, \psi$ known we may then quantify the interference pattern and locate the particle.

To make clear the contribution of this work, we claim that to determine the height of the measurement plane, the particle position and the phase shift it is sufficient to simply measure the length of two major axes. If the measurement height is known then the particle position and phase shift can be determined from a single measurement. To clarify these statements we now proceed to find the position of the particle which  generates the intereference pattern of Figure \ref{BigFringe}.

Figure \ref{BigFringe} represents an image taken over an area of 60$\times$60$ \mu$m. In Table \ref{Positions} we give the measured values of the major and minor axis lengths as well as the position of the centre (in the $(X,Y)$ system) for the first four brightest fringes. To minimise errors this data was obtained in the following manner: first identify the top and bottom points of the first  ellipse, which we denote $(X_t, Y_t), (X_b, Y_b)$. Since the fringes are of finite width we take a central point within each fringe which appears to be in the brightest region: the distance between the two points is $2b_n$. This is repeated three times and then averaged, to obtain the $b_n$ presented in Table \ref{Positions}. The same process is repeated to calculate $a_n$. The whole process is then repeated for the surrounding fringes.
The centre point was obtained by averaging the co-ordinates used to find $a_n, b_n$.

\begin{table}[h!]
\centering
\begin{tabular}{ |c|c|c|c| }
 \hline
   & $a_n$ & $b_n$ & $(X_c, Y_c)$\\
  \hline
 Fringe 1 & 7.29& 12.33  & (38.94, 26.78)   \\
 Fringe 2 & 11.60& 19.40  & (37.85, 27.62) \\
 Fringe 3 & 14.97& 24.46  & (37.51, 28.33) \\
 Fringe 4 & 17.38 & 29.00  & (38.10, 29.28) \\
 \hline
\end{tabular}
\caption{Length of minor and major axes and position of centre of the fringes.  All lengths are in $\mu$m}
\label{Positions}
\end{table}

Below we will use only the first two fringes, which are the clearest in the figure.
Calculating $\theta_i = \cos^{-1}(a_n/b_n)$ for these two fringes indicates an average value $\theta_i = 53.52^{\circ}$ which is very close to the quoted value of $53^{\circ}$.
Now we numerically solve the following equations
\bea
b_n &=& 12.33 = \frac{\sqrt{\Lambda_n^2-z_c^2\cos^2 \theta_i}}{\cos^2\theta_i}\\
b_{n+1} &=& 19.40 = \frac{\sqrt{\Lambda_{n+1}^2-z_c^2\cos^2 \theta_i}}{\cos^2\theta_i}
\eea
to find $z_c = 36.06$ $\mu$m $\mu_n = 43.32$ $\mu$m. Writing  $\mu_n = C \lambda$, so $C = 43.32/0.632 \approx 68.55$, we immediately deduce $n=68$, $\psi = 2\pi \times 0.55 \approx 3.45$.

To compare with the experimental image requires converting between the two co-ordinate systems.
The $\mathbf{x}$ system has been chosen so that the particle is located at the origin and the $y$-axis aligned with the major axis of the brightest ellipse. The conversion  therefore requires a rotation and translation.
When measuring $b_n$ we recorded the top and bottom points of the first two fringes (in the $\mathbf{X}$ system). This permits us to calculate the rotation angle between the $x$ and $X$ axes, $ \tan \theta = (X_t-X_b)/(Y_t-Y_b) \approx -4.42^{\circ}$. This indicates that to move from the $\mathbf{x}$ system to the $\mathbf{X}$ system requires a rotation of $\theta_r =4.42^{\circ}$ in the clockwise direction.
The appropriate transformation is then
\bea
\label{CSeq}
\left( \begin{array}{c}  X\\
 Y \\ Z   \end{array} \right)
=\left( \begin{array}{c}  X_p\\
 Y_p \\ 0   \end{array} \right) +
\left( \begin{array}{c c c} \cos\theta_r & -\sin \theta_r & 0 \\ \sin \theta_r & \cos \theta_r & 0 \\
 0 & 0 & 1 \end{array} \right) \left( \begin{array}{c}  x\\
 y \\ z   \end{array} \right) ~ ,~~~~
\eea
where $(X_p, Y_p, 0)$ is the unknown position of the particle in the $\mathbf{X}$ system.
According to equation \eqref{ycneq} the centre of the first visible  fringe is  at
\bea
y_{c(68)} = (\mu_{68}-z_c \cos\theta_i)\frac{\sin\theta_i}{\cos^2\theta_i} \approx 49.76 ~ .
\eea
 The position $(x_{c(68)},y_{c(68)})=(0, 49.76)$ must coincide with the centre quoted  in Table \ref{Positions}, $(X_{c(68)},Y_{c(68)}) = (38.94, 26.78)$. Using these two co-ordinates in equation \eqref{CSeq} determines the location of the particle in the $\mathbf{X}$ system
\bea
\mathbf{X_p} &=& (X_{c(68)} + y_{c(68)} \sin \theta_r, Y_{c(68)} - y_{c(68)} \cos \theta_r,0) \nonumber \\
 & \approx& (42.43, -22.83,0) ~ .
\eea

\begin{figure}[htb!]
	\begin{center}
	\includegraphics[width=0.5\textwidth]{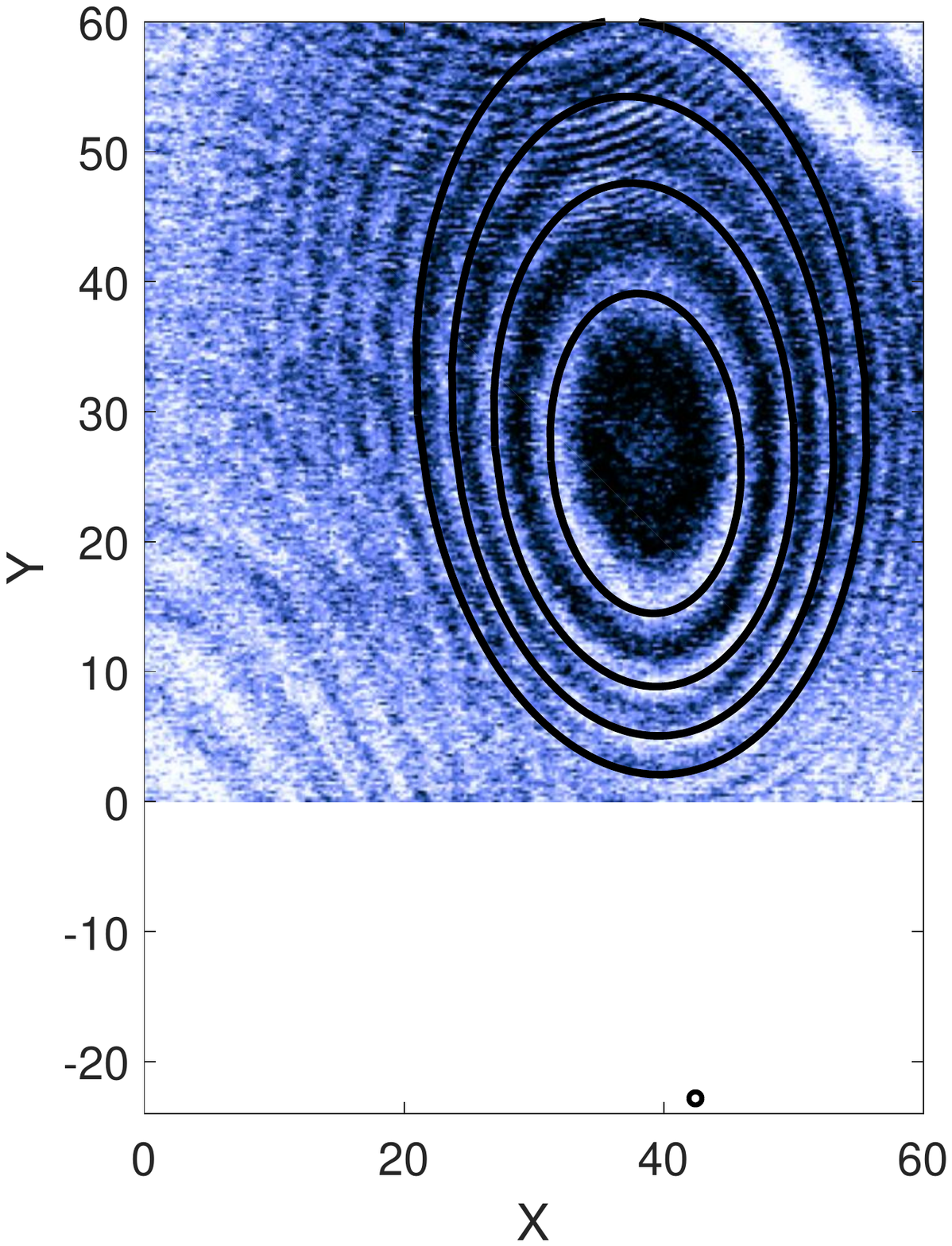}\vspace{-3.0cm}
	\end{center}	
	\caption{Comparison of experimental and theoretical curves, black lines represent the theoretical prediction. The circle at (42.43, -22.83) represents the nanoparticle. }
	\label{AnalFringe}
\end{figure}

Using the above transformation on the ellipse equation, \eqref{Ellipseeq}, the first four theoretical fringes are plotted alongside the experimental ones in Figure \ref{AnalFringe}. Clearly there is excellent correspondence between theory and experiment. The circle at (42.43, -22.83) represents the position of the nanoparticle. A further check on the accuracy comes from substituting the predicted values of $z_c, n, \psi$ into the expressions for  $b_{n+2}, b_{n+3}$ which gives $b_{69} = 24.64, b_{70} = 29.06$ $\mu$m, these are within 0.7\% of the measured values shown in Table \ref{Positions}. However, whilst the values are all excellent, it may be observed that beyond the first fringe the theoretical curves show a slight shift in the positive $X$ direction, which increases with increasing $n$. There are a number of possible reasons for this discrepancy, such as errors in the determination of the fringe position, errors in the piezo scanner along the $x$ and $y$ direction and finite size effects of the nanoparticle (as the measurement plane approaches the substrate the approximation of a point scatterer will become less accurate).

An important feature of this solution method is the ease with which the phase shift is obtained. In the past this has been the subject of complex retrieval methods, see the overview given in Shechtman {\it et al.} \cite{Shecht}. Here we obtain a value by simply measuring two distances. To understand why it is generally considered such a complex problem  it is worth considering the issue of sensitivity.

The phase shift occurs in the governing equation for the fringes, equation \eqref{parabeq}, through the parameter  $\mu_n = n \lambda(1+\epsilon)$ where $\epsilon = \psi/(2n\pi)$. Since $\psi \in [0, 2\pi)$ the value $\epsilon = \ord{1/n}$ and so $\epsilon \ll 1$ for reasonably large $n$.
For the current study where $n=68$ the value of $\epsilon$ is of the order $1/68 \approx 0.015$. This means, for example, that a 1.5\% error in the measurement of $b_n$ could completely change the predicted value of $\psi$. Put another way, the value of $n, z_c$ are relatively insensitive to the $b_n$ measurement: an error of the order 1\% results in errors of the order 1\% in $n, z_c$ but possible errors of the order 100\% in $\psi$. Equation \eqref{parabeq} describes the surfaces of maximum brightness, so the sensitivity is inherent to the system and not a result of our current experimental set-up (which at the moment only deals with horizontal cross-sections). The only way to decrease sensitivity is to decrease $n$, i.e. take measurements closer to the particle but this introduces another error. The theoretical model is based on the assumption that the particle is a point source, as we move closer to it this approximation becomes less accurate. In order to improve accuracy we must seek some balance between proximity to the particle and the point source approximation.

In summary, the comparison with experiment demonstrates that with only two simple measurements the theory can accurately reproduce the fringe pattern. Equation (\ref{parabeq}) shows the sensitivity of various parameters to measurement errors. Importantly, from these measurements we are able to accurately locate the nanoparticle. However, in accordance with existing studies on phase retrieval, the shift $\psi$ is sensitive to measurement errors. The sensitivity  decreases as the fibre approaches the particle but at the same time the approximation that the particle is a point becomes less realistic.
For this reason, the present method is a simple way to locate nanoparticles. To accurately determine the phase shift requires further research, for example into improving the quality of the images and image analysis and also ascertaining optimal heights for the measurement.

TM and HR acknowledge funding for Short Term Scientific Missions to CEMES through COST Action, TD1409, Mathematics for Industry Network. TM acknowledges the support of Ministerio de Ciencia e Innovacion Grant No. MTM2017-82317-P and also CNRS-INP. HR and WB acknowledge funding grant NEXT No. ANR-17-EURE-0009 in the framework of the Programme des Investissements d’Avenir. We thank F. Neumayer and S. Weber for experimental support.

\bibliographystyle{natbib}

\end{document}